\documentclass[aps,amsmath,amssymb,reprint, double column,prl, showkeys,superscriptaddress]{revtex4-2}

\usepackage{graphicx,epsfig}
\usepackage{amssymb}
\usepackage{amsmath}
\usepackage{bm}
\usepackage{textcomp}
\usepackage{color}
\usepackage{float}
\usepackage{comment}

\usepackage{soul}

\begin{document}
\setcounter{page}{1}

\title[]{Signatures of correlated defects in an ultra-clean Wigner crystal in the extreme quantum limit}
\author{P. T. \surname{Madathil}}
\affiliation{Department of Electrical and Computer Engineering, Princeton University, Princeton, New Jersey 08544, USA}
\author{C. \surname{Wang}}
\affiliation{Department of Electrical and Computer Engineering, Princeton University, Princeton, New Jersey 08544, USA}
\author{S. K. \surname{Singh}}
\affiliation{Department of Electrical and Computer Engineering, Princeton University, Princeton, New Jersey 08544, USA}
\author{A. \surname{Gupta}}
\affiliation{Department of Electrical and Computer Engineering, Princeton University, Princeton, New Jersey 08544, USA}
\author{K. A. \surname{Villegas Rosales}}
\affiliation{Department of Electrical and Computer Engineering, Princeton University, Princeton, New Jersey 08544, USA}
\author{Y. J. \surname{Chung}}
\affiliation{Department of Electrical and Computer Engineering, Princeton University, Princeton, New Jersey 08544, USA}
\author{K. W. \surname{West}}
\affiliation{Department of Electrical and Computer Engineering, Princeton University, Princeton, New Jersey 08544, USA}
\author{K. W. \surname{Baldwin}}
\affiliation{Department of Electrical and Computer Engineering, Princeton University, Princeton, New Jersey 08544, USA}
\author{L. N. \surname{Pfeiffer}}
\affiliation{Department of Electrical and Computer Engineering, Princeton University, Princeton, New Jersey 08544, USA}
\author{L. W. \surname{Engel}}
\affiliation{National High Magnetic Field Laboratory, Florida State University, Tallahassee, Florida 32310, USA}
\author{M. \surname{Shayegan}}
\affiliation{Department of Electrical and Computer Engineering, Princeton University, Princeton, New Jersey 08544, USA}

\date{\today}

\begin{abstract}

Low-disorder two-dimensional electron systems in the presence of a strong, perpendicular magnetic field terminate at very small Landau level filling factors in a Wigner crystal (WC), where the electrons form an ordered array to minimize the Coulomb repulsion. The nature of this exotic, many-body, quantum phase is yet to be fully understood and experimentally revealed. Here we probe one of WC’s most fundamental parameters, namely the energy gap that determines its low-temperature conductivity, in record-mobility, ultra-high-purity, two-dimensional electrons confined to GaAs quantum wells. The WC domains in these samples contain $\simeq 1000$ electrons. The measured gaps are a factor of three larger than previously reported for lower quality samples, and agree remarkably well with values predicted for the lowest-energy, intrinsic, hyper-corelated bubble defects in a WC made of flux-electron composite fermions, rather than bare electrons. The agreement is particularly noteworthy, given that the calculations are done for disorder-free composite fermion WCs, and there are no adjustable parameters. The results reflect the exceptionally high quality of the samples, and suggest that composite fermion WCs are indeed more stable compared to their electron counterparts.

\end{abstract}

\maketitle  
 
In sufficiently dilute two-dimensional electron systems (2DESs), when the Coulomb energy is much larger than the kinetic energy, electrons should arrange themselves into an ordered array called the Wigner crystal (WC) \cite{wigner1934interaction}. Quantitatively, a quantum WC is expected when the parameter $r_s$, defined as the ratio of the Coulomb and kinetic energies exceeds $\simeq$35 in a 2DES \cite{tanatar1989ground,attaccalite2002correlation,drummond2009phase}. Since the theoretical proposal of Wigner, several experiments in various platforms have reported signatures of a 2D WC \cite{grimes1979evidence,yoon1999crystallization,hossain2020observation,zhou2021bilayer,smolenski2021signatures,shayegan2022wigner,falson2022competing,hossain2022anisotropic}. While reducing the density in an ideal 2DES eventually leads to Wigner crystallization, the disorder present in real 2DESs makes it extremely challenging to realize the WC state \cite{shayegan2022wigner}. Another way to quench the kinetic energy and induce the WC phase is to subject the 2DES to a large, perpendicular magnetic field which quantizes the electron energy to discrete, flat Landau levels. In high quality 2DESs, a magnetic-field-induced WC phase is generally stabilized in the lowest Landau level at small filling factors ($\nu$) $\lesssim$ 1/5. Several early experiments reported evidence for the WC phase at high-fields \cite{shayegan1997perspectives,shayegan2022wigner}. These included non-linear \textit{I-V} \cite{willett1989current,goldman1990evidence,williams1991conduction,jiang1991magnetotransport}, noise \cite{li1991low,li1995rf,li1996inductive}, and microwave resonance \cite{andrei1988observation, ye2002correlation, chen2006melting} measurements. More recently, some intricate properties of the magnetic-field-induced WC were probed in NMR \cite{tiemann2014nmr}, bilayer commensurability \cite{deng2016commensurability}, and tunneling \cite{jang2017sharp} experiments. However, some of most basic and fundamental characteristics of this exotic, quantum state of matter are yet to be established.

A fundamental parameter that is intimately linked to the formation and defect energies of the WC is the activation energy ($E_A$) that determines its low-temperature conduction. Numerical calculations indicate that the formation energy of the high-field WC is minimized when the crystal makes use of the composite fermion correlations, as opposed to the traditional, Hartree-Fock electron crystal \cite{yi1998laughlin,narevich2001hamiltonian,chang2005microscopic,archer2013competing}. However, the experimental $E_A$, determined from the temperature dependence of the sample resistance at very low fillings \cite{willett1988termination,jiang1991magnetotransport,paalanen1992electrical}, were about 10 times smaller than the theoretical predictions. Later calculations, that considered the energies associated with \textit{defects} in a composite fermion WC, shown in Fig. 1(b) \cite{archer2014quantum}, reduced the discrepancy between theory and experiments \cite{footnote.relevant_defect}, but the theoretical defect energy still remained a factor of $\sim$3 larger than the experimental values. 
 
 On the experimental front, the ubiquitous disorder present in 2DESs masks the intrinsic energy gap and contributes significantly to a reduction of $E_A$. Disorder causes the WC to lose long-range correlations and break into domains \cite{fisher1979defects,cha1994orientational,ruzin1992pinning,cha1994topological}. With less disorder, larger domains can be realized, allowing for a closer comparison between theory and experiments. A recent breakthrough \cite{chung2021ultra,chung2022understanding} in the quality (purity) of dilute 2DESs confined to modulation-doped GaAs quantum wells has resulted in an improvement of the 2DES mobility by more than a factor of 10 over previous samples \cite{willett1988termination,jiang1991magnetotransport,paalanen1992electrical}, enabling us to probe the physics of more ideal (less disordered) WCs at very low fillings. We report here $E_A$ measurements made on these ultra-pure 2DESs. The data show a remarkable quantitative agreement with theoretical calculations, suggesting the presence of a composite fermion WC with non-classical, exotic defects.


\begin{figure*}[t]
\centering
\includegraphics[width=1\textwidth]{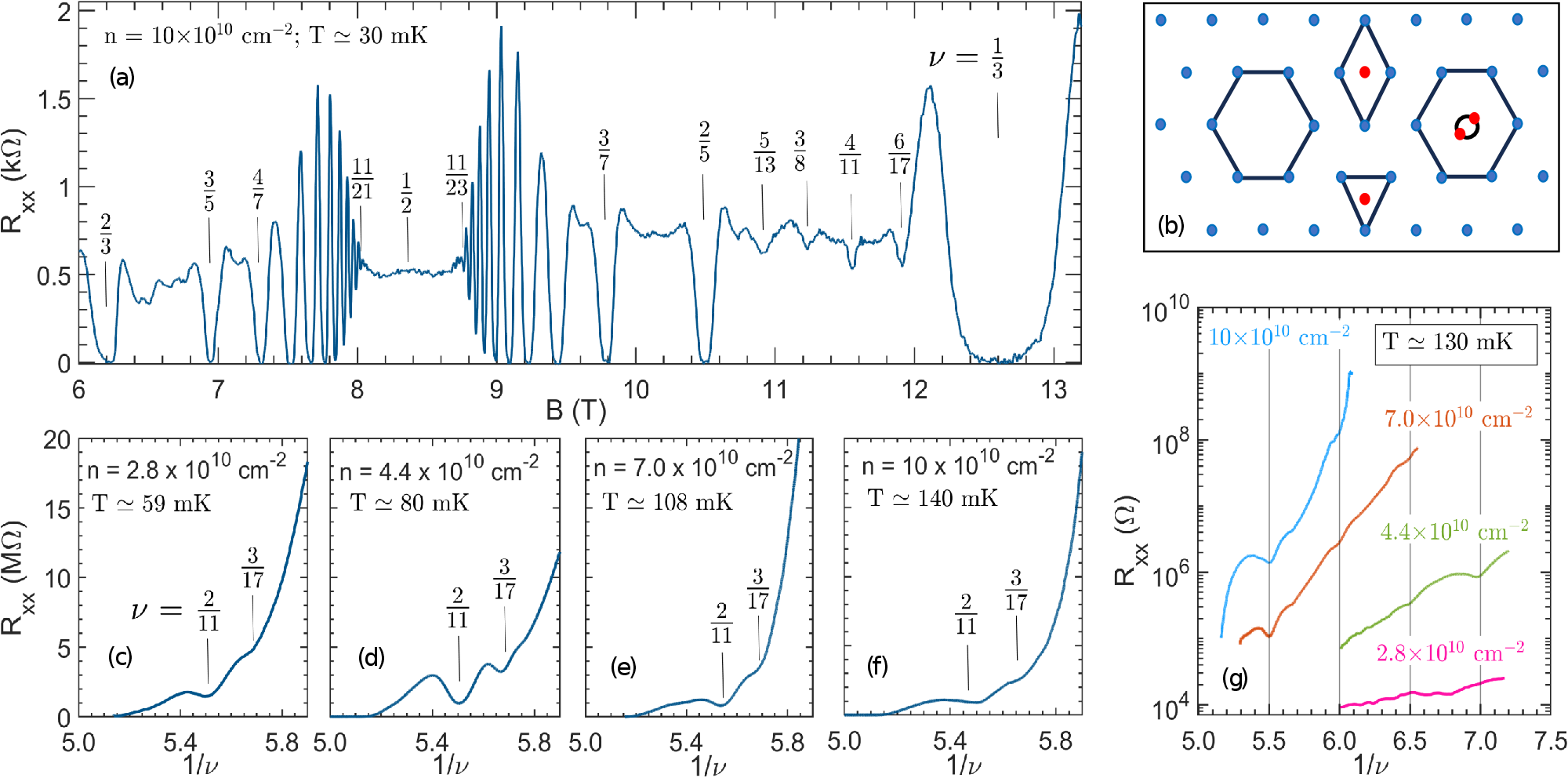}
\centering
  \caption{\label{Ip} 
(a) Longitudinal resistance ($R_{xx}$) vs. magnetic field ($B$) for a 2DES of density 10$\times 10^{10}$ cm$^{-2}$, confined to a 50-nm-wide GaAs quantum well at $T\simeq$ 30 mK in the filling range ($2/3 > \nu > 1/3)$. (b) Different defect configurations of a composite fermion WC considered in calculations by Archer \textit{et al.} \cite{archer2014quantum}. A vacancy is shown on far left, edge and center interstitials in the center (top and bottom), and the hyper-correlated bubble defect on the far right. (c-f) $R_{xx}$ vs. $1/\nu$ for four different densities. The positions of emerging FQHSs at $\nu = 2/11$ and 3/17 are marked. Note that the scales are the same in all four panels, but the traces were taken at different temperatures, as indicated. (g) $R_{xx}$ vs. $1/\nu$ for the four different densities in a semi-log plot at $T\simeq$ 130 mK.
  }
  \label{fig:Ip}
\end{figure*}

We performed activation energy measurements in four different ultra-pure samples with densities $n$ = 2.8, 4.4, 7.0 and 10, in units of $10^{10}$ cm$^{-2}$ which we use throughout the manuscript. The extremely high quality 2DESs are realized by confining electrons to a GaAs quantum well bounded by Al$_x$Ga$_{1-x}$As stepped barriers of  varying $x$, and placing Si dopants inside a doping-well structure on either side of the quantum well \cite{chung2021ultra,chung2022understanding}. The samples' well widths are 89, 70, 58, and 50 nm, respectively, and their mobilities range from $\simeq 15\times 10^6$ for the lowest density, to $\simeq 36\times 10^6$  cm$^2$/Vs  for the highest density sample. We performed electrical transport measurements in the van der Pauw geometry, with alloyed In:Sn serving as contacts at the corners and center edges of 4$\times$4 mm$^2$ samples. The samples were cooled in a dilution refrigerator and transport experiments were carried out using low-frequency, lock-in techniques.

\begin{figure*}[t]
\centering
\includegraphics[width=0.98\textwidth]{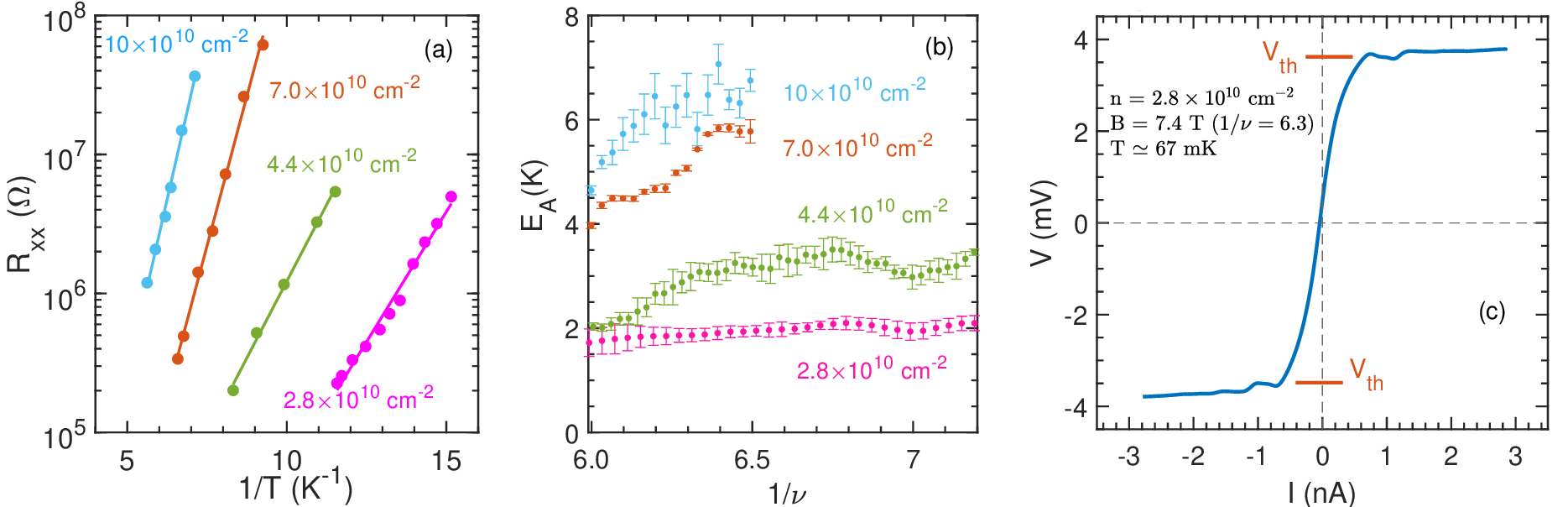}
\centering
  \caption{\label{Ip} 
(a) $R_{xx}$ at $\nu = 1/6$ vs. 1/$T$ shown in an Arrhenius plot. The slope of each fits gives the activation energy gap ($E_A)$. (b) $E_A$ vs. 1/$\nu$ for the four different densities in the filling factor range of 6.0 $\leq 1/\nu \leq$ 7.2. (c) Current-voltage ($I-V$) data for the $n=2.8$ sample at $B$ = 7.4 T (1/$\nu = 6.3$) and $T \simeq$ 67 mK.
  }
  \label{fig:Ip}
\end{figure*}

Figure 1(a) displays a representative trace of the longitudinal resistance ($R_{xx}$) vs. magnetic field ($B$) for a 2DES with $n$ = 10, measured at $T\simeq30$ mK. The exceptional quality of the 2DES is evident from the rich features displayed in the filling range $2/3 >\nu >$ 1/3. We see fractional quantum Hall states (FQHSs) up to $\nu$ = 11/23 and 11/21, corresponding to the standard Jain sequence for 2-flux CFs on the electron and hole side of $\nu$ = 1/2, respectively \cite{jain2007composite}. We also observe emergent unconventional FQHSs between $\nu$ = 1/3 and 2/5, at filling factors $\nu$ = 3/8, 4/11, 5/13 and 6/17. These very fragile states between the conventional Jain states can be viewed as FQHSs of \textit{interacting} CFs \cite{wojs2004fractional,mukherjee2014enigmatic}, and are seen only in the highest quality samples \cite{chung2021ultra,chung2022correlated,pan2015fractional,samkharadze2015observation}.

Figures 1(c-f) show a part of the high-field traces for the four samples between the inverse filling factor (1/$\nu$) values of 5.0 and 5.9. Note that the full $R_{xx}$ scale is 20 M$\Omega$, four orders of magnitude larger than in the low-field trace shown in Fig. 1(a). The 2DESs in the range $1/\nu >5.2$ exhibit a strongly insulating behavior characteristic of the high-field WC phases at very small fillings \cite{shayegan2022wigner,shayegan1997perspectives,willett1989current,goldman1990evidence,williams1991conduction,jiang1991magnetotransport,li1991low,li1995rf,li1996inductive,andrei1988observation,ye2002correlation,chen2006melting}. To measure resistances of the order of 10 M$\Omega$ and larger, we used a pre-amplifier (SR551) with an input impedance $\gtrsim$ 1 T$\Omega$. The driving $ac$ current (0.05 nA) and frequency (0.2 Hz) were kept sufficiently small to avoid non-linear effects, and to maximize the in-phase (resistive) component of the signal. In Figs. 1(c-f) traces, we see a clear minimum at $\nu = 2/11$ and a minimum or an inflection at $\nu = 3/17$, suggesting competing FQHSs amidst the strong, insulating WC background. Emerging FQHSs deep in the insulating regime, competing against the insulating, WC solid phase have been reported for $1/5 > \nu > 1/6$, and at even a smaller filling of $\nu = 1/7$ in very high quality samples \cite{pan2002transition,chung2022correlated,goldman1988evidence}. Note that, in the older samples where $E_A$ was measured \cite{willett1988termination,jiang1991magnetotransport}, none of the high-field FQHSs past $\nu = 1/5$ were reported, further attesting to the ultra-high quality of our new 2DESs.

The resistance scale is kept uniform in Figs. 1(c-f). As seen in these figures, the temperatures at which similar magnitude resistances are attained for the same $\nu$ are very different for the four samples. Figure 1(g) shows $R_{xx}$, presented in a log-scale vs. $1/\nu$, for different densities at $T$ $\simeq$ 130 mK. Inflections and local minima, corresponding to the Jain FQHSs at $\nu = 2/11$, 3/17, 2/13, and 1/7 can be seen in some of the traces. Near $\nu = 1/6$, where we have data for all four densities, $R_{xx}$ changes by over four orders of magnitude from the lowest density to the highest, indicative of the extremely insulating nature of the samples at these fillings.

Figure 2(a) displays representative Arrhenius activation plots for our samples. The activation energy, $E_A$ is determined from the temperature dependence of $R_{xx}$, given by the relation $R_{xx} \propto e^{E_A/2k_BT}$, where $k_B$ is the Boltzmann constant. The data shown in Fig. 2(a) are obtained at $\nu = 1/6$.
The data show excellent agreement with the expected activated behavior over a resistance change of about two decades for each density. As we move to higher densities, the slope clearly increases, reflecting a higher activation energy. We repeat this analysis, and obtain $E_A$ for the four different densities in a range of small fillings as shown in Fig. 2(b).

\begin{figure}[t]
\centering
\includegraphics[width=0.48\textwidth]{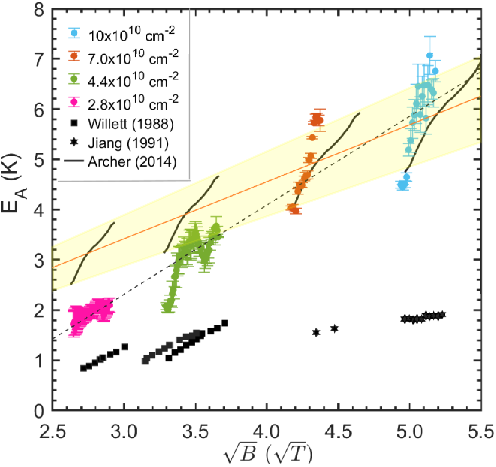}
\centering
  \caption{\label{Ip} 
Activation energy ($E_A$) vs. $\sqrt{B}$ for experimental and theoretical activation data. The circles are our experimental data points extracted from measurements on ultra-pure samples. The dotted line is a least-squares-fit through all the experimental data points. The squares and stars represent previous experimental data reported in Refs. \cite{willett1988termination} and \cite{jiang1991magnetotransport}, respectively. The solid curves represent theoretical values of the CFWC defect energy $E_D$ reported in Ref. \cite{archer2014quantum}. The yellow band highlights the range of the calculated $E_A$ in the filling range 6.0 $\leq1/\nu \leq$ 7.2 for each density, and the orange solid line represents $E_A$ calculated for $1/\nu = 6.5$.
  }
  \label{fig:Ip}
\end{figure}

Figures 2(b) and 3 contain our main findings. We summarize the extracted $E_A$ for all four samples as a function of 1/$\nu$ (Fig. 2(b)) and $\sqrt{B}$ (Fig. 3). In both figures, it is clear that the energy gaps are generally larger at higher densities. The relevant energy scale in the high-field WC is the Coulomb energy given by the expression, $E_C = e^2/4\pi\epsilon \epsilon_{0} l_B$, where $l_B = \sqrt{\hbar/eB}$. In an ideal WC, $E_A$ (at a fixed filling) would then be expected to have a $\sqrt{B}$ dependence. Our experimental data overall scale reasonably well with $\sqrt{B}$, as seen from the dotted line in Fig. 3 which is a least-squares fit through all the data points. However, there are deviations from the expected behavior, which we will address later in the manuscript.

Before discussing our measured $E_A$ quantitatively, we make a few remarks regarding the disorder in our samples. The high-field WC phase is generally considered to be broken into domains because of the ubiquitous disorder present in the 2DES \cite{fisher1979defects,ruzin1992pinning,cha1994orientational,cha1994topological,chitra1998dynamical}. The domain size provides a measure of the correlation length of the WC and its long-range order. Very recent microwave resonance experiments on samples of quality similar to those we use in our study report an estimated domain size of $\sim$ 1000 electrons per domain for $n=4.4$ \cite{lili2023dynamic}. This estimate is corroborated by our non-linear current-voltage \textit{I-V} measurements \cite{madathilunpublished}. As shown in Fig. 2(c), for our sample with $n=2.8$, we observe a voltage threshold of $E_{th}\simeq 4$ mV at $1/\nu \simeq 6.3$, corresponding to an electric field threshold of $E_{th}\simeq$ 1 V/m. Using the expression $L_0^2\simeq$ $0.02e/4\pi\epsilon\epsilon_0E_{th}$ \cite{fukuyama1978dynamics}, we find the number of electrons in a domain, $nL_0^2$ $\sim$ 600. Since the number of electrons in a domain should scale with density, we estimate that the domains contain $\sim$ 2000 electrons for our highest-density sample. This is approximately 13 times larger than in previously reported samples \cite{willett1988termination,jiang1991magnetotransport, williams1991conduction, ye2002correlation,chen2006melting}. In Ref. \cite{jiang1991magnetotransport}, e.g., the threshold electric field for the onset of non-linear \textit{I-V} was $\simeq 16$ V/m at $1/\nu \simeq 6.3$ for a
sample with $n = 11$, implying $\sim$ 150 electrons in a typical WC domain. The much larger domains in our samples render them more ideal for quantitative comparisons with defect energy models of the intrinsic WCs that do not contain disorder.

The magnitude of the measured activation energies presented in Fig. 3 are indeed remarkably consistent with theoretical predictions made for the \textit{intrinsic} defect formation energies ($E_D$) of an exotic WC. Several calculations in the clean limit indicate that the WC state at high magnetic fields is a \textit{composite fermion} WC (CFWC) rather than a Hartree-Fock \textit{electron} crystal \cite{yi1998laughlin,narevich2001hamiltonian,chang2005microscopic,archer2013competing}. The 2DES takes advantage of the CF correlations to minimize the energy and form a crystal. Archer \textit{et al.} \cite{archer2014quantum} also considered the intrinsic defects in an otherwise pristine CFWC, and calculated their $E_D$. The lowest $E_D$ should correspond to the transport gaps measured in experiments \cite{footnote.ED}. Archer \textit{et al.} \cite{archer2014quantum} found that a unique, non-classical, hyper-correlated bubble defect (see the defect configuration in the far right of Fig. 1(b)) has the lowest $E_D$.

In Fig. 3 we provide a comparison between our measured $E_A$ values and $E_D$ calculated in Ref. \cite{archer2014quantum} for the hyper-correlated bubble defect. The four theoretical curves in Fig. 3 are for the four densities of our samples, and are based on calculated $E_D$ in the filling factor range $1/6 \leq \nu \leq 1/7.2$. . These correspond to 4-flux CFWCs with an additional vortex attached to the bubble defect. The yellow “band” in Fig. 3 highlights the range of the calculated $E_D$ for $1/6 \leq \nu \leq 1/7.2$ \cite{footnote.ED}. Note that, at a given $\nu$, the calculated $E_D$ follows a $\sqrt{B}$ dependence and its center (along the y-axis) follows the $\sqrt{B}$ dependence and intercepts the axes at the origin, as expected for an ideal system with no disorder (see orange line in Fig. 3). The overall agreement between theory and experimental gaps is impressive, especially considering the fact that there are no adjustable parameters in making the comparison. Particularly noteworthy is the fact that the experimental data closely follow the calculations that are performed for the clean limit. This reflects the exceptionally high quality of the new samples, and is consistent with the extremely large WC domain sizes which are estimated to contain $\sim$ 1000 electrons (or CFs).

It is worth emphasizing that theoretical studies favor the CFWC over the (Hartree-Fock) electron WC \cite{yi1998laughlin,narevich2001hamiltonian,chang2005microscopic,archer2013competing}. Moreover, the lowest defect-formation energy predicted for the Hartree-Fock WC is about a factor of 3 larger than that of the CFWC \cite{archer2014quantum}. Quantitatively, our experimentally measured $E_A$ values are clearly in much better agreement with the calculations for a CFWC. The observation of developing FQHSs at $\nu = 2/11$, 3/17, 2/13, and 1/7, which can be understood as the integer quantum Hall states of 6-flux CFs, also demonstrates the presence of CFs and supports the formation of a CFWC near these fillings.

In our data, we do find deviations from theory for the high-density \cite{footnote.dev} and low-density samples. In the low-density case, disorder might be a culprit. It has been theoretically shown that disorder plays a significant role in lowering $E_A$ \cite{cha1994orientational,cha1994topological,chui1991finite}. In Fig. 3, we also show previous experimental data reported in Refs. \cite{willett1988termination} and \cite{jiang1991magnetotransport}. The activation energies are lower by a factor of 2 to 3 compared to our new ultra-pure 2DESs and theoretical calculations that do not include disorder. This can be attributed to the much greater disorder and the ensuing smaller WC domains in the old samples, leading to large deviations in $E_A$ from what one would expect in the disorder-free limit. In Fig. 3, the smaller $E_A$ we measure in our low-density samples, compared to $E_A$ predicted for the clean limit, can also be related to a reduction of the domain size in our most dilute samples.

Archer \textit{et al.} \cite{archer2014quantum} also report a reduction in $E_A$ due to finite (non-zero) layer thickness of the 2DES. Our lowest-density samples have the largest well widths (89 and 70 nm), and would therefore be most affected. (To minimize interface roughness and inter-subband scattering, our samples are fabricated with as large of a GaAs well width as possible, while ensuring that the upper electric subband is not occupied \cite{chung2021ultra, chung2022understanding}). Lastly, the calculations also ignore the role of Landau level mixing. Since Landau level mixing scales as 1/$\sqrt{B}$, one would again expect the largest reduction in $E_A$ for the most dilute samples. It is possible that more rigorous calculations that include finite layer thickness and Landau level mixing quantitatively would lead to an even better agreement with the measured gaps.

In summary, we report energy gap measurements on several ultra-pure, dilute 2DESs, in the very small Landau level filling ($\nu< 1/5$) range, where the 2DES is deep in the insulating phase that is the hallmark of the field-induced WC. Thanks to the world-record-high purity of our samples, the WC domains are $\sim$ 10 times larger than in previous samples, and contain $\sim$ 1,000 electrons, rendering the new samples much closer to the ideal, zero-disorder case. Our measured gaps are about a factor of three larger than those previously reported for lower quality samples. The new gaps agree remarkably well with calculations for the lowest-energy, intrinsic, correlated defects in a WC made of flux-electron CFs, rather than defects in a Hartree-Fock electron crystal. The observation of developing FQHSs in the vicinity of the WC crystal fillings also demonstrates the formation of CFs, and provides further evidence that the crystal is composed of CFs as opposed to bare electrons. Finally, we remark that, while new experimental studies are shedding light on the periodic structure of the WC, both at zero field \cite{shayegan2022wigner} and high magnetic fields \cite{Tsui2023direct}, the energy gaps reported here provide a very fundamental, macroscopic parameter which is intimately linked to the microscopic details of a quantum WC, namely its quasi-particle ingredients and intrinsic defects.

We acknowledge support by the U.S. Department of Energy Basic Energy Office of Science, Basic Energy Sciences (Grant No. DEFG02-00-ER45841) for measurements. For sample characterization, we acknowledge support by the National Science Foundation (NSF) Grant No. DMR 2104771 and, for sample synthesis, NSF Grants No. ECCS 1906253 and the Gordon and Betty Moore Foundation’s EPiQS Initiative (Grant No. GBMF9615 to L.N.P.). This research is funded in part by a QuantEmX travel grant from ICAM and the Gordon and Betty Moore Foundation through Grant GBMF9616. A portion of this work was performed at the National High Magnetic Field Laboratory (NHMFL), which is supported by the NSF Cooperative Agreement No. DMR-1644779 and the state of Florida. We thank A. Bangura,  D. Graf, G. Jones, T. Murphy, and R. Nowell at NHMFL for technical support. We also thank J. K. Jain for illuminating discussions.



\begin{thebibliography}{99}
\bibitem{footnote.dev} For the two high-density samples, the measured $E_A$ exhibit a faster rise with $B$ compared to the calculations. While we do not fully understand this behavior, we remark that at the highest fields, $R_{xx}$ for these samples reach extremely high values (exceeding $\sim 10^7 \Omega$). This limits the temperature range over which $R_{xx}$ exhibits an activated behavior, making it difficult to extract accurate values for $E_A$.
\end{thebibliography}

\end{document}